\documentstyle[preprint,tighten,floats,aps,epsf]{revtex}

\begin{document}
\title{Exotic shapes in $^{40}$Ca and $^{36}$Ar studied with Antisymmetrized
Molecular Dynamics}

\author{Y. Kanada-En'yo, M. Kimura$^*$ and H. Horiuchi$^{**}$}

\address{{Institute of Particle and Nuclear Studies, 
High Energy Accelerator Research Organization, Tsukuba 305-0801, Japan\\
*The Institute of Physical and Chemical Research, Wako 351-0198, Japan\\
** Department of Physics, Kyoto University, Kyoto 606-01, Japan\\}}

\maketitle
\begin{abstract}
The structures of the ground and excited states of $sd$-shell nuclei:
$^{40}$Ca and $^{36}$Ar were studied with a method of antisymmetrized 
molecular dynamics. Recently observed rotational bands were described
with deformed intrinsic states which are dominated by excited configurations
such as $4p-4h$, $8p-8h$ and $4p-8h$ states. 
The results showed the coexistence of 
various kinds of exotic shapes with oblate, prolate and large prolate
deformatios in the low-energy region of $^{40}$Ca.
Possible cluster aspects in 
heavy $sd$-shell nuclei were implied in these exotic shapes.
\end{abstract}

\noindent

\section{Introduction}

Owing to the experiments of gamma-ray measurements,
many excited bands in the nuclei near $^{40}$Ca have 
been recently observed \cite{IDEGUCHI,SVENSSON}.
The rotational bands with large moments of inertia are 
hot subjects relating with super deformations in $sd$-shell
nuclei. In $^{40}$Ca, there exist
many low-lying bands, which imply the coexistence of various shapes.
It is interesting problem why the shape coexistence
occurs even in the doubly magic nucleus.
The shape coexistence problem and the mechanism of deformations 
in these $sd$-shell nuclei is one of the attractive subjects.

 Our interests are possible exotic shapes and cluster aspects
in the deformed excited states.
It has been already known that 
clustering is one of the essential features in light nuclei .
Exotic shapes are formed due to cluster structures in some light nuclei, 
For example, a famous cluster structure in $^{12}$C is three $\alpha$ clusters which 
form a triangle shape. Parity doublets of the rotational bands in
$^{20}$Ne are described by a parity asymmetry shape with 
a $^{16}$O+$\alpha$ cluster structure.
It is a long problem whether or not cluster aspects appear 
in heavier nuclei. 
Many of the familier cluster structures are those in light stable nuclei.
In the recent experimental and theoretical 
research \cite{FREER,SAITO,ENYOg,ITAGAKI,ENYObe14}, 
cluster structures were suggested in light unstable
nuclei such as $^{10}$Be and $^{12}$Be, 
where new type clusters such as $^{6}$He and $^{8}$He are proposed.
Our aim in this paper is systematic study of the
$sd$-shell nuclei near $^{40}$Ca
to find possible exotic states with cluster aspects.

For the systematic study of ground and excited states of
$sd$-shell nuclei, theoretical difficulties
exist in the coexistence of cluster and mean-field aspects.
Needless to say, cluster aspects are important in light nuclei, 
while the mean-field is an essential feature in heavy nuclei. 
Since the $sd$-shell is an inter-mediate mass-number region, where
both aspects should be significant,
we need a miscroscopic method beyond the traditional clsuter models.
Namely, we adopt a method of antisymmetrized molecular dynamics(AMD)
\cite{ONOa,ENYOa,ENYObc,ENYOe,ENYOg}.
The method AMD was applied for light nuclei and has been proved
to be a powerful approach to describe the ground and excited states of 
unstable nuclei as well as stable nuclei. 
In the pioneering study with the AMD method,
the structure changes on the ylast line of $^{20}$Ne are explained
in terms of alteration of the cluster structure \cite{ENYOa}.
In the recent studies of light unstable nuclei with AMD, 
a new cluster aspect in unstable nuclei was suggested 
that cluster cores develop in the deformed neutron mean-field in
neutron-rich nuclei.

In the present work, we adopt a new effective force with a finite-rage 
three-body term, with which 
the binding energies and the radii of $^{4}$He, $^{16}$O and
$^{40}$Ca are systematically reproduced, 
because the usual effective forces used in the AMD study,
such as Minnesota forces, Volkov forces 
and Modified Volkov forces, have serious problems in
reproducing these basic properties in the wide mass-number region. 

In this paper, the structures of the ground and excited states of 
$^{40}$Ca and $^{36}$Ar are studied by the AMD method. We investigate
the shape coexistence problem of these nuclei. The 
mechanism of deformations are discussed, focusing on
cluster aspects. We also introduce the AMD study of neuton-rich 
$sd$-shell nuclei near $^{32}$Mg.

\section{Formulation}
 \label{sec:formulation}

In this section, the formulation of AMD 
for the nuclear structure study of ground and excited states
is briefly explained. For more detailed descriptions of the AMD framework,
the reader is referred to Refs \cite{ENYOa,ENYObc,ENYOe}.

The wave function of a system is written by AMD wave functions,
\begin{equation}
\Phi=c \Phi_{AMD} +c' \Phi '_{AMD} + \cdots .
\end{equation}
An AMD wave function of a nucleus with a mass number $A$
is a Slater determinant of Gaussian wave packets;
\begin{eqnarray}
&\Phi_{AMD}({\bf Z})=\frac{1}{\sqrt{A!}}
{\cal A}\{\varphi_1,\varphi_2,\cdots,\varphi_A\},\\
&\varphi_i=\phi_{{\bf X}_i}\chi_{\xi_i}\tau_i :\left\lbrace
\begin{array}{l}
\phi_{{\bf X}_i}({\bf r}_j) \propto
\exp\left 
[-\nu\biggl({\bf r}_j-\frac{{\bf X}_i}{\sqrt{\nu}}\biggr)^2\right],\\
\chi_{\xi_i}=
\left(\begin{array}{l}
{1\over 2}+\xi_{i}\\
{1\over 2}-\xi_{i}
\end{array}\right),
\end{array}\right. 
\end{eqnarray}
where the $i$th single-particle wave function $\varphi_i$
is a product of the spatial wave function $\phi_{{\rm X}_i}$,
 the intrinsic spin function $\chi_{\xi_i}$ and 
the iso-spin function $\tau_i$. 
The spatial part $\phi_{{\rm X}_i}$ is presented by 
variational complex parameters $X_{1i}$, $X_{2i}$, $X_{3i}$.
$\chi_{\xi_i}$ is the intrinsic spin function defined by
$\xi_{i}$, and $\tau_i$ is the iso-spin
function which is fixed to be up(proton) or down(neutron)
in the present calculations.
The values 
${\bf Z}\equiv \{X_{ni},\xi_i\}\ (n=1,2,3\ \hbox{and }  i=1,\cdots,A)$
are the variational parameters which express an AMD wave function.

  In order to obtain the wave functions of ground and excited states,
we perform a generator coordinate method in the framework of AMD.
First we vary the energy of the parity eigen state projected from an
AMD wave function under a constraint that the total oscillator quanta
 must equal to a given
number ${\cal N}$ as $\langle a a^\dagger \rangle={\cal N}$.
The energy variation is numerically calculated by a frictional cooling 
method \cite{ENYOa}. An energy curve is obtained as a function of the 
coordinate ${\cal N}$ in the constraint.
In the second step, the spin-parity eigen states projected from 
the obtained AMD wave functions are superposed by 
diagonalizing Hamiltonian and Norm matrices,
$\langle P^{J\pm}_{MK'}\Phi_{AMD}({\bf Z}_i)|H|P^{J\pm}_{MK''}\Phi_{AMD}
({\bf Z}_j)\rangle$ and
$\langle P^{J\pm}_{MK'}\Phi_{AMD}({\bf Z}_i)|P^{J\pm}_{MK''}\Phi_{AMD}
({\bf Z}_j)\rangle$,
We call the present calculations as VBP(variation before projection) 
because the variation is performed before the spin projection.  

\section{Interactions} 
\label{sec:interaction}

The ordinary effective forces with no three-body term
such as the Volkov force or with a zero-range three-body term like the 
MV1 force are not appropriate to describe the binding energies and radii of
nuclei covering wide mass number region from $\alpha$ to $^{40}$Ca.
Hence we use an interaction containing a finite-range three-body term in
the present calculations.
The central part of the interaction  is explained by
a conbination of the two-body and three-body terms.
The interaction parameters used in the present paper are as follows, 
\begin{eqnarray}
&V_{central}=\sum_{i<j}V^{(2)}+\sum_{i<j<k}V^{(3)},\\
&V^{(2)}=(1-m+bP_\sigma-hP_\tau-m P_\sigma P_\tau)
\left\lbrace 
V_a \exp[-({r_{12}\over r_a})^2]
V_b \exp[-({r_{12}\over r_b})^2]\right\rbrace \\
&+V_c \exp[-({r_{12}\over r_c})^2],\\
&V^{(3)}=V_d \exp[-d(r_{12}^2+r_{23}^2+r_{31}^2)^2],\\
&V_a=-198.34\ {\rm MeV}, V_b=300.86\ {\rm MeV}, V_c=22.5\ {\rm MeV}, \\ 
& r_a=1.2\ {\rm fm}, r_b=0.7\ {\rm fm} , r_c=0.9\ {\rm fm},V_d=600\ {\rm MeV}, d=0.8\ {\rm fm}^{-2}\\
& m=0.193, b=-0.185, h=0.37,
\end{eqnarray}
where the width and strength parameters are chosen so as to reproduce 
reasonably the sizes of $\alpha$ and $^{40}$Ca, and the binding energies of 
$\alpha$, $^{16}$O and $^{40}$Ca.
In choosing parameters, $\alpha$+$\alpha$ phase shifts and 
the saturation property of symmetric nuclear matter have been also 
taken into consideration. 
The adopted interaction is 
a sum of the central force, the G3RS spin-orbit force \cite{LS}
with the strength $u_{ls}=2500$ MeV and the Coulomb force.

\section{Results}\label{sec:results}

\begin{figure}\caption{\label{fig:exp}
The experimental data of the energy levels of $^{40}$Ca.
The figure is taken from \protect\cite{IDEGUCHI}.
}
\epsfxsize 12cm\centerline{\epsffile{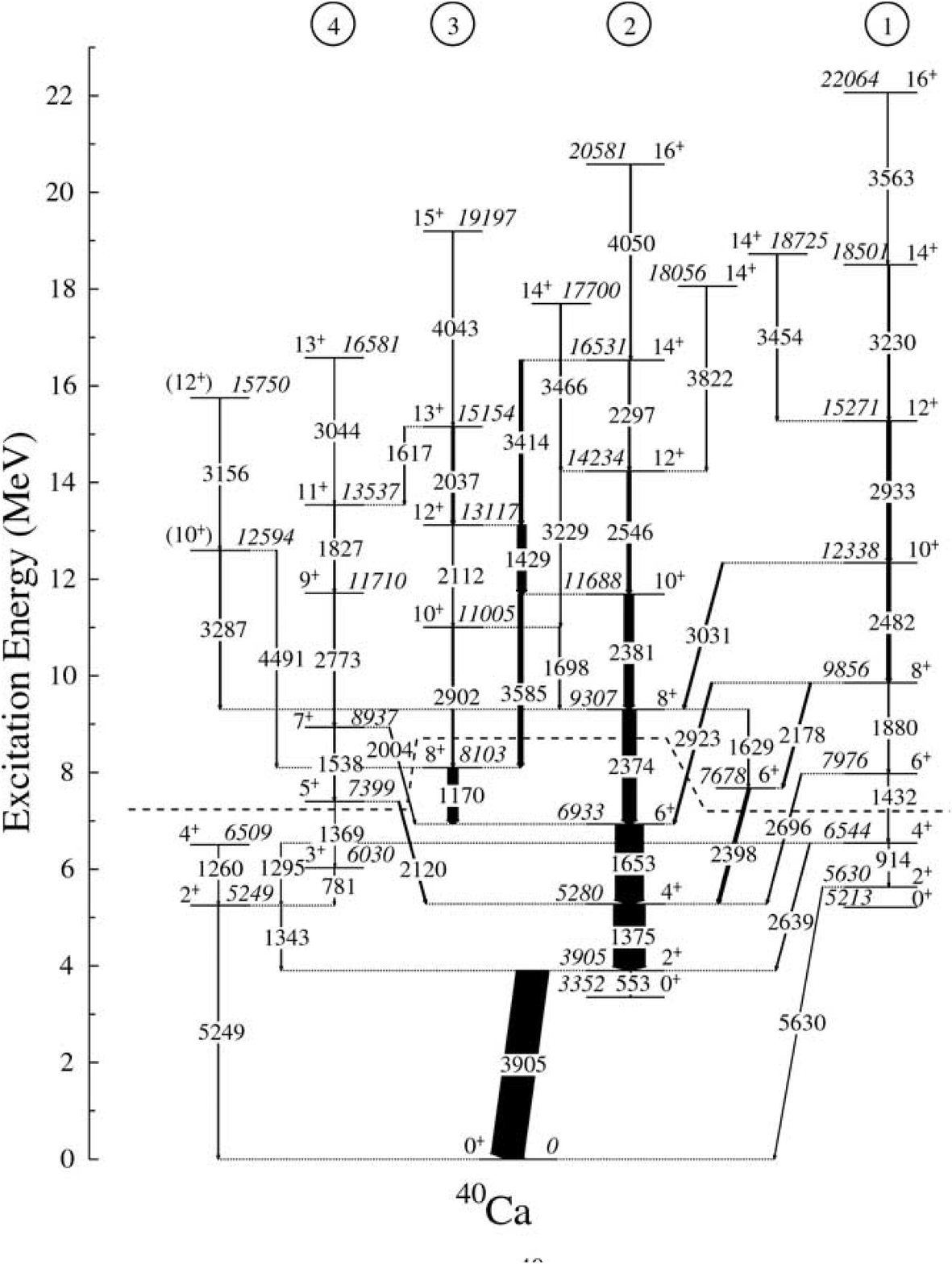}}
\end{figure}

The excited deformed bands in the $^{40}$Ca and $^{36}$Ar have 
been recently observed in the 
gamma-ray measurements \cite{IDEGUCHI,SVENSSON}, which reveal
the existence of many low-lying bands in these nuclei.
In the experimental level scheme of $^{40}$Ca,
the rotational band with a large moment of inertia (No.1 in
Fig.\ref{fig:exp})
is a hot topic relating with a possible super deformation.

\begin{figure}\caption{\label{fig:level}
The theoretical energy levels of $^{40}$Ca obtained after the diagonalization
with respect to the spin-parity projected states.}
\epsfxsize 15cm\centerline{\epsffile{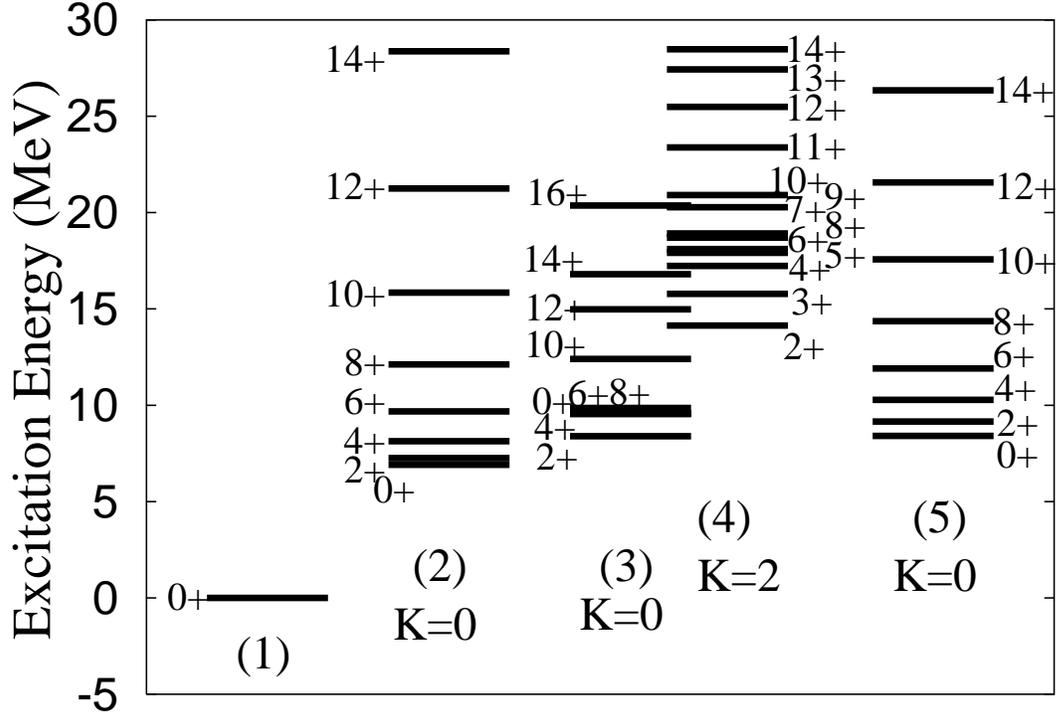}}
\end{figure}

We study the ground and excited states 
of $^{40}$Ca and $^{36}$Ar,
with the method of a generator coordinate method in the framework of 
AMD with respect to the total oscillator quanta.
In the energy curves obtained after spin-projection as a function of the
oscillator quanta, we can not necessarily find local minima except for
the absolute minimum.
However, after diagonalization of the Hamiltonian
and Norm matrices, the energy curves for the excited states have other
local minima due to the conditions orthogonal to the lower states.
As a result, we obtain many low-liying
rotational bands (Fig.\ref{fig:level}), each of which is dominated
by the spin-parity projected states from an intrinsic AMD wave function.

By analyzing deformations of these dominant intrinsic states, 
it is found that various shapes coexist in $^{40}$Ca.
The interesting point is that each of the intrinsic states 
in $^{40}$Ca is well described by either of $0p-0h$, $4p-4h$ or $8p-8h$
configurations.
First we obtain the normal spherical ground state
with the doubly closed $sd$-shell configuration. With the 
increase of the oscillator quanta, the deformed excited 
states appear in the low-energy region. In the region concerning $4p-4h$
states, there exist opposite shapes as two local minimum states.
One is an oblate shape with the deformation parameter $\beta=0.2$, 
while the other is a prolate deformation with $\beta=0.2$.
The oblate state provides the excited $K^\pi=0^+$ band (2),
and the bands (3) and (4) with $K^\pi=0^+, 2^+$ are dominated by
the states projected from the prolate intrinsic state.
The reason for the higher $0^+$ state than the $2^+$ state
in the $K^\pi=0^+$ band (3) is the mixing with the ground state.
With a larger oscillator quanta, we find another largely prolate
deformation which is dominated by $8p-8h$ configurations.
The energy levels of the band (5) obtained from this 
large deformation indicate a large moment of inertia of this
rotational band.
It is very surprising that the excited band with such highly excited 
configurations as $8p-8h$ exists in low-energy region.
We identify this band as the experimentally observed super-deformed band
which starts from the $0^+$ state at 5.2 MeV.

As shown in Table \ref{tab:be2}, 
the theoretical values of the $E2$ transition strength
in the band (5) are extremely large comparing
with those in other excited band.
According to a quantitative comparison of the strength in the band (5) 
with the experimental data,
the theoretical $E2$ strength for $4^+\rightarrow 2^+$ is smaller than
the corresponding experimental data
$B(E2;6.54{\rm MeV} \rightarrow 5.63{\rm MeV})=100 (W.u.)$.
Therefore, it is conjectured that the intrinsic deformation $\beta=0.4$ 
of the band (5) in the present results must be still small 
compared with the value $\beta=0.6$ for a 
typical super deformation.
The underestimation of the deformations in the present theory will be 
improved by the superposition along generator coordinates and also the 
extension of the single-particle wave functions.

\begin{table}
\caption{ \label{tab:be2} The theoretical values (W.u.) of $E2$ transition 
strength in the excited $K^\pi=0^+$ bands.}
\begin{center}
\begin{tabular}{cccc}
 transition & band(2) & band(3) & band(5)\\ 
\hline
 $2^+\rightarrow 0^+$ & 11 & 2 & 30  \\
 $4^+\rightarrow 2^+$ & 16 & 21 & 53  \\
 $6^+\rightarrow 4^+$ & 16 & 13 & 61  \\
 $8^+\rightarrow 6^+$ & 17 & 14 & 78  \\
\end{tabular}
\end{center}
\end{table}

\begin{figure}\caption{\label{fig:density}
The density distributions of the intrinsic states before 
spin-parity projections. The notations (2),(3),(5) are same as those in
Fig.\protect\ref{fig:level}. In each figure, the density of the
dominant AMD wave function of each band is presented.
The matter density is integrated along the 
axis with the largest moment of inertia.}
\epsfxsize 13cm \centerline{\epsffile{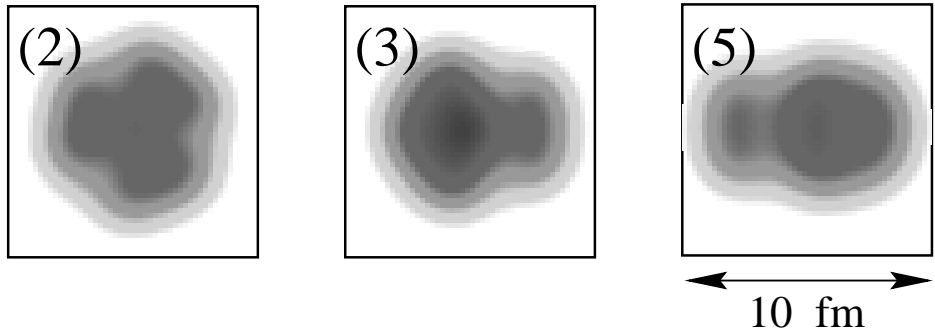}}
\end{figure}

As mentioned above, the point in the present results of $^{40}$Ca
is the coexistence of various shapes: the spherical state, the
oblate shape, the normal prolate deformation and the large prolate
deformation.
By analysing the intrinsic structures of these deformed states, 
exotic shapes and cluster aspects are revealed.
The density distributions of the dominant intrinsic states of the 
excited bands are presented in Fig.\ref{fig:density}. 
As shown in Fig.\ref{fig:density},
the oblate state (2) has a hexagon structure like a three-leaf clover
which consists of three $^{12}$C cores surrounding 
an $\alpha$ at the center. 
In the super deformation, the system has a parity-asymmetry shape like a
pear because of a $^{28}$Si+$^{12}$C-like clustering.
We focus the characteristics of the $^{28}$Si+$^{12}$C-like
cluster structures in the super deformation.
The first point is that the clusters are not weak coupling
 but are strong coupling ones. 
It is interesting that the $^{12}$C cluster consists of four 
nucleons in the $sd$-shell and eight nucleons in the $pf$-shell. Namely, it is a
inter-shell cluster which lies over different shells. 
Because of the strong coupling feature, it is expected that 
the deformed mean-field effect can be important as well as cluster
aspects in the state.
In second, if such a cluster structure indeed develops, it forms a parity
asymmetric deformation, which may provide the parity doublets made of positive
and negative-parity bands.
Thirdly, it is reasonable that both clusters are the sub-shell
closed nuclei. In other words, the stability of these cluster cores 
is based on the shell effects of the sub-shell closure.
For further analysis of the cluster aspects, we need
advanced calculations where the residual effects such as the 
coupling of other cluster channels, the tail of inter-cluster motion, and the
core deformations are more carefully taken into consideration.

It should be pointed that the present results imply
the possible exotic shapes due to the cluster effects 
such as $^{12}$C cores, though there are few experimental evidences 
of the existence of clusters in $^{40}$Ca.
The investigation of the level structures of the negative-parity states 
is one of the probes, because 
the parity-doublets of positive and negative-parity bands 
support a cluster structure as the origin of 
a parity-asymmetry intrinsic state.
The present calculations predict a
$K^\pi=0^-$ band with the band-head $1^-$ state at 11.7 MeV
, which is about 3 MeV higher 
than theoretical band-head energy of the positive-parity super deformation.
The deformation of the dominant intrinsic state is rather large as
$\beta=0.3$ and the intrinsic structure is similar to that of the
super deformation in the positive-parity states.
Therefore we consider this negative-parity band
as a candidate of the parity doublet.
Strictly speaking, in negative parity
states, the rotational levels fragments because the state mixing 
between bands is rather strong.
As a result, the intra-band $E2$ transitions in the $K^\pi=0^-$ band
are not so strong as those in the positive-parity band of the
super deformation.

In the theoretical results of $^{36}$Ar, we find 
excited bands with a large prolate deformation $\beta=0.3$ 
which are dominated  by $4p-8h$ configurations. This band starts from 
a 4 MeV excitation energy, and well corresponds to the newly 
observed rotational 
band with a large moment of inertia in the experimental data \cite{SVENSSON}.

At the end of this section, I briefly report the AMD study 
on neutron-rich $sd$-shell
nuclei near $^{32}$Mg calculated by one of the authors(M. K.).
The structures of the ground and excited states of $^{32}$Mg are hot 
subjects concerning the island of inversion of a neutron magic number
$N=20$. 
The ground and excited states of Mg isotopes are 
calculated by an extended version of AMD with the Gogny force.
The calculations well reproduce the energy levels of $^{32}$Me 
and the features of the island of inversion. 
In the neutron-rich $sd$-shell nuclei, the deformed neutron mean-field 
plays an important role. The unique predictions in this work are 
the low-lying negative parity states in $^{32}$Mg.
Since the nuclei $^{32}$Mg and $^{40}$Ca have the same neutron
number, it is interesting future problem to compare the intruder states
of $^{32}$Mg with the excited states of $^{40}$Ca.

\section{Summary}
The structures of the ground and excited states of $sd$-shell nuclei:
 $^{40}$Ca and $^{36}$Ar were studied with the method of antisymmetrized 
molecular dynamics. The recently observed excited bands were well described
with the deformed intrinsic states in the present calculations.
It was found that various kinds of shape coexist in the low-energy
region. In $^{40}$Ca, it was suggested in the present results that
the oblate shape, 
the normal prolate deformation and the large prolate deformation
coexist in the low-lying excited bands 
besides the spherical ground state.
The oblate, normal prolate, and the large prolate states are 
explained by the excited configurations with
$4p-4h$, $4p-4h$, and $8p-8h$, respectively.
It is a mystery that the rotational band with the highly excited 
configurations($8p-8h$) starts from the low-energy region.
The exotic shapes due to the cluster effects were implied
in the intrinsic structures of these deformed states.
The present results suggested possible cluster aspects in heavy $sd$-shell
nuclei.

\acknowledgments
Authors would like to thank Professors A. Tohsaki, Y. Akaishi
and K. Ikeda for helpful discussions and comments.
They are also thankful to Dr. N. Itagaki and E. Ideguchi 
for many discussions. 
This work was partially performed in the ``Research Project for Study of
Unstable Nuclei from Nuclear Cluster Aspects'' sponsored by
Institute of Physical and Chemical Research (RIKEN).
The computational calculations of this work are supported by 
the Supercomputer Project No.58, No.70 of High Energy Accelerator Research
Organization(KEK), and Research Center for Nuclear Physics(RCNP) 
in Osaka University.


\end{document}